\documentclass[aps,pra,preprint,superscriptaddress,floatfix,longbibliography]{revtex4-1}

\usepackage{amsmath}    
\usepackage{amssymb}
\usepackage{graphicx}   
\usepackage[lofdepth,lotdepth, caption=false]{subfig}
\usepackage{verbatim}   
\usepackage{color}      
\usepackage{hyperref}   
\usepackage{dcolumn}
\usepackage{xcolor}  

\begin{document}
\newcommand{\sit}{Sn$_{1-x}$In$_{x}$Te}
\newcommand{\psit}{(Pb$_{1-x}$Sn$_{x}$)$_{1-y}$In$_{y}$Te}
\newcommand{\psitn}{(Pb$_{0.5}$Sn$_{0.5}$)$_{0.7}$In$_{0.3}$Te}
\newcommand{\pst}{Pb$_{1-x}$Sn$_{x}$Te}
\newcommand{\pstn}{Pb$_{0.5}$Sn$_{0.5}$Te}
\newcommand{\degrees}{$^\circ$C}
\newcommand{\Tc}{$T_{c}$}

\title{Indium substitution effect on the topological crystalline insulator family (Pb$_{1-x}$Sn$_{x}$)$_{1-y}$In$_{y}$Te: Topological and superconducting properties}

\author{Ruidan~Zhong}
\email{rzhong@bnl.gov}
\affiliation{Condensed Matter Physics and Materials Science Department, Brookhaven National Laboratory, Upton, NY 11973, USA}
\affiliation{Materials Science and Engineering Department, Stony Brook University, Stony Brook, NY 11794, USA}
\author{Qiang~Li}
\affiliation{Condensed Matter Physics and Materials Science Department, Brookhaven National Laboratory, Upton, NY 11973, USA}
\author{Wei~Ku}
\affiliation{Condensed Matter Physics and Materials Science Department, Brookhaven National Laboratory, Upton, NY 11973, USA}
\affiliation{Department of Physics and Astronomy, Shanghai Jiao Tong University, Shanghai 200240, China}
\affiliation{Tsung-Dao Lee Institute, Shanghai 200240, China}
\author{J.~M.~Tranquada}
\affiliation{Condensed Matter Physics and Materials Science Department, Brookhaven National Laboratory, Upton, NY 11973, USA}
\email{jtran@bnl.gov}
\author{Genda Gu}
\affiliation{Condensed Matter Physics and Materials Science Department, Brookhaven National Laboratory, Upton, NY 11973, USA}

\date{\today} 

\begin{abstract}
Topological crystalline insulators (TCIs) have been of great interest in the area of condensed matter physics. We investigated the effect of indium substitution on the crystal structure and transport properties in the TCI system \psit.  For samples with a tin concentration $x\le50\%$, the low-temperature resisitivities show a dramatic variation as a function of indium concentration: with up to $\sim2\%$ indium doping the samples show weak-metallic behavior, similar to their parent compounds; with $\sim6\%$ indium doping, samples have true bulk-insulating resistivity and present evidence for nontrivial topological surface states; with higher indium doping levels, superconductivity was observed, with a transition temperature, \Tc, positively correlated to the indium concentration and reaching as high as 4.7~K. We address this issue from the view of bulk electronic structure modified by the indium-induced impurity level that pins the Fermi level. The current work summarizes the indium substitution effect on (Pb,Sn)Te, and discusses the topological and superconducting aspects, which can be provide guidance for future studies on this and related systems.
\end{abstract}

\maketitle
\section{Introduction}
Topological insulators (TIs) are a class of materials that are currently the focus of considerable attention since they represent a new state of matter in which the bulk is an insulator with an "inverted" energy gap induced by a strong spin-orbit coupling (SOC), which leads to the emergence of unusual gapless edge or surface states protected by time-reversal-symmetry\cite{Hasan2010, Qi2011, Ando2013, Kane2011}. First discovered in two-dimensional systems, one of the simplest topological insulators is the quantum spin Hall state, in which the SOC plays the same role as the magnetic field in the quantum Hall effect. In the quantum Hall effect, the bulk conductance is zero while the edge states are conducting with current flowing in one direction around the edge of the system. 
Similarly, in the quantum-spin-Hall state, the bulk is still insulating while edge-state electrons with opposite spins propagate in opposite directions, consistent with time-reversal symmetry. 
Theoretical concepts were soon generalized to three dimensions and shown experimentally in materials such as Bi$_{1-x}$Sb$_{x}$\cite{Fu2007}. As in the 2D case, the direction of an electron's motion along the surface of a 3D topological insulator is locked to the spin direction, which now changes continuously as a function of propagation direction, resulting in an unusual "planar metal". In the bulk of a TI, the electronic band structure resembles that of an ordinary band insulator, with the Fermi level falling between the conduction and valence bands. On the surface of a TI there are special states that fall within the bulk energy gap and allow surface metallic conduction. Although ordinary band insulators can also support conductive surface states: the locking of the spin and propagation directions eliminates the possibility of backscattering from nonmagnetic impurities.

The first key experiment in this field was the observation of the 2D quantum-spin-Hall effect in a quantum-well structure made by sandwiching a thin layer of mercury telluride (HgTe) between layers of mercury cadmium telluride (Hg$_{x}$Cd$_{1-x}$Te), following a theoretical prediction\cite{Bernevig2006}. The first 3D TI to be probed using angle-resolved photoemission spectroscopy (ARPES) was the semiconducting alloy Bi$_{1-x}$Sb$_{x}$\cite{Hsieh2008}. Simpler versions of the 3D TI were theoretically predicted in Bi$_{2}$Te$_{3}$, Sb$_{2}$Te$_{3}$\cite{Zhang2009} and Bi$_{2}$Se$_{3}$\cite{Zhang2009, Xia2009} compounds with a large bulk gap and a gapless surface state consisting of a single Dirac cone. Later ARPES experiments indeed observed the linear dispersion relation of these surface states\cite{Xia2009, Chen2009}. These discoveries confirmed the ubiquitous existence in nature of this new topological state. 

In 2011, the notion of "topological crystalline insulators (TCIs)" was introduced to extend the topological classification of band structures to include certain crystal point group symmetries\cite{Fu2011}. This new state of matter features metallic surface states with quadratic band dispersion on high symmetry crystal surfaces, and it was shown that such a situation is realized in an insulating crystal having rocksalt structure. It has caused quite a sensation since the first example, SnTe, has been theoretically\cite{Hsieh2012} and experimentally\cite{Tanaka2012} confirmed to exhibit topological surface states in <001>, <110> and <111> surfaces. Soon after this discovery, the topological surface state in the Pb-doped \pst\ and Pb$_{1-x}$Sn$_{x}$Se have been verified by ARPES and Landau level spectroscopy using scanning tunneling microscopy and spectroscopy\cite{Dziawa2012, Xu2012, Yan2014}, thus expanding the range of relevant materials. Alongside SnTe and the related alloys Pb$_{1-x}$Sn$_{x}$Se/Te, other chalcogenides such as SnS and SnSe that incorporate lighter elements have also been predicted to be TCIs even without the SOC\cite{Sun2013}. In theory, by applying external pressure, normal IV-VI rocksalt chalcogenides can be tuned into TCIs\cite{Barone2013}. Besides the materials with rocksalt crystal structure, Hsieh $et.al.$ predicted that the antiperovskite family are also promising materials for exploring the topological and other related properties\cite{Hsieh2014}. More recently, a new phase of Bi is stablized by strain, has been found to be a TCI based on mirror symmetry, similar to SnTe\cite{Munoz2016}. 

The discovery of TIs and TCIs has also stimulated the search for topological superconductors (TSCs), whose surfaces should exhibit Majorana fermions\cite{Qi2011}. Superconductors derived from TIs by doping have been considered as TSC candidates, such as Cu$_{x}$Bi$_{2}$Se$_{3}$\cite{Fu2010odd, Hor2010sc}. Since the topological surface states are protected from backscattering by disorder, it should be safe to tune the chemical potential through chemical substitution.
The ARPES studies performed on \sit\ (SIT) at $x$ = 0.045 confirmed that the topological surface states remain intact after In doping\cite{Sato2013}. In fact, SnTe becomes a superconductor upon substituting 2\% or more of Sn with In, which introduces hole carriers\cite{Erickson2009, Balakrishnan2013, Zhong2013, Novak2013}. Similarly, doping the TCI \pstn\ with > 10\% indium also induces superconductivity\cite{Zhong2014}. This spurs interest in searching for the superconducting analogue, a time-reversal-invariant topological superconductor (TSC), in this system. Point-contact spectroscopy experiments performed on SIT with varies In concentrations found that a zero-bias conductance peak is observed only in the cleanest samples with $x \approx$ 0.04, suggesting that there is competition between topological and non-topological superconducting states, and that disorder may determine the outcome\cite{Novak2013}. 

A challenge for characterizing the transport properties of surface states
in TI/TCI materials such as Bi$_{2}$Se$_{3}$ and SnTe is the dominance of a pronounced bulk conductance\cite{Skinner2012, Butch2010}. Despite considerable efforts to reduce the bulk carrier density, such as modifying crystal growth method\cite{Jia2011}, reducing sample thickness\cite{Peng2009} and chemical counterdoping\cite{Hor2009, Analytis2010, Checkelsky2011}, the bulk conduction has proved difficult to suppress. Inspired by the goal of finding truly bulk-insulating topological materials, we have found that indium doping the TCI materials (Pb,Sn)Te can yield a huge bulk resistivity while maintaining topological surface states\cite{Zhong2015}. 

In this article, we present a review of the effects of indium substitution on the crystal structure, resistivity behavior, and electronic band structure in the TCI family \psit\ (PSIT). By varying the indium concentration, samples show an extreme range of low-temperature resistivities: with a few percent indium doping, the samples show weak-metallic behavior; with $\sim$6\% indium doping, samples have true bulk-insulating resistivity and present evidence for nontrivial topological surface states; with higher indium doping levels, superconductivity with a transition temperature \Tc\ positively correlated to the indium concentration was observed. We consider this behavior from the standpoint of the localized impurity states associated with the indium dopants. 

\section{Results}
\subsection{Crystal structure}

\begin{figure}
\centering
\includegraphics[width=13cm]{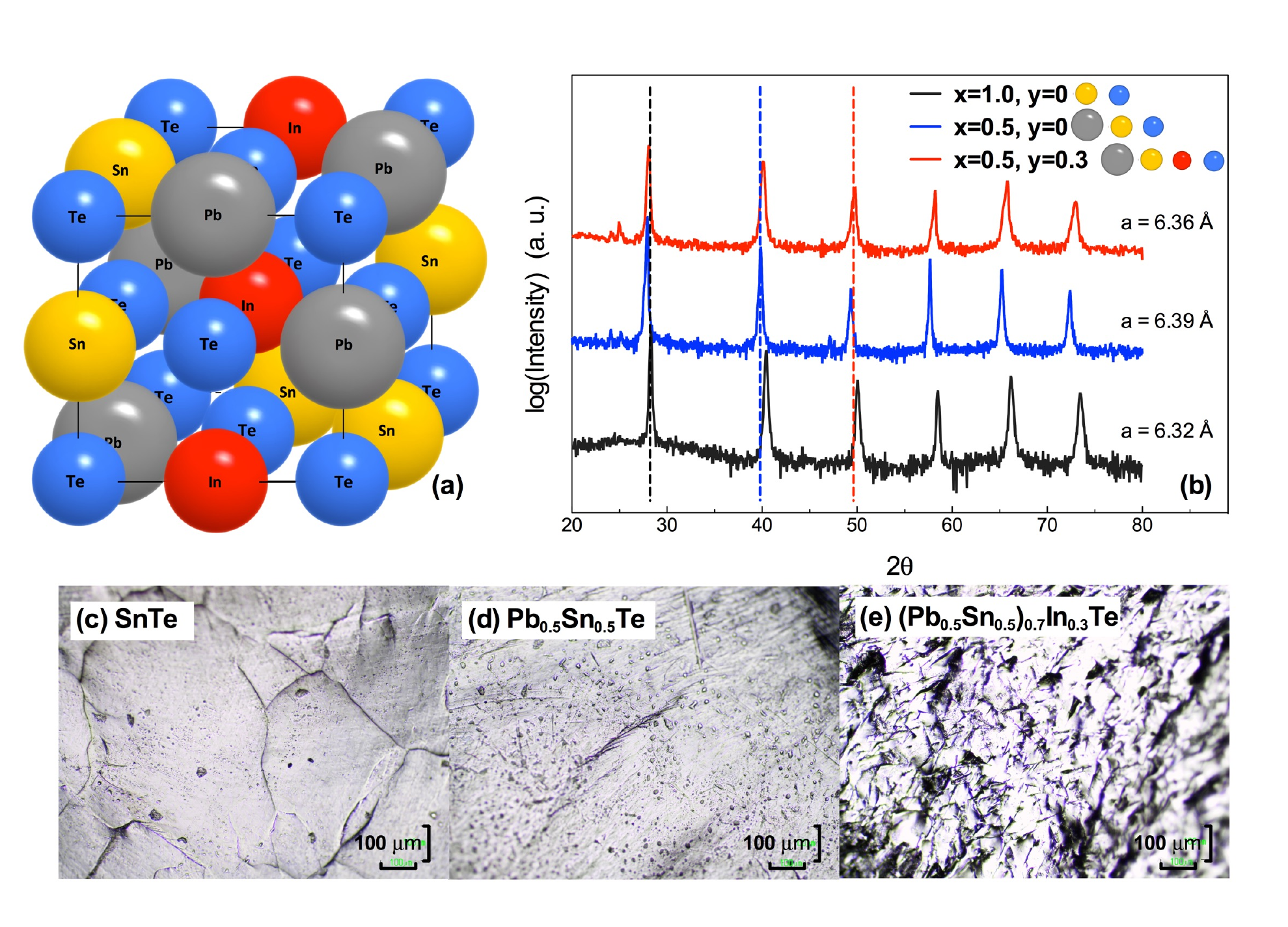}
\caption{\label{fig:crystal} (color online) (\textbf{a}) A sketch of the crystal structure of SnTe with Sn atoms (yellow) partially replaced by Pb (grey) and In (red). (\textbf{b}) X-ray powder diffraction (XRD) patterns for SnTe (black), \pstn\ (blue) and (Pb$_{0.5}$Sn$_{0.5}$)$_{0.7}$In$_{0.3}$Te (red), respectively. Each dashed line marks the position of an XRD peak of a compound with the same color. (\textbf{c-e}) Optical microscope photos of the pristine surface of SnTe (c), \pstn\ (d) and (Pb$_{0.5}$Sn$_{0.5}$)$_{0.7}$In$_{0.3}$Te (e).}
\end{figure}   

SnTe is a IV-VI semiconductor that crystallizes in the cubic rocksalt structure at room temperature, and maintains this structure after a certain degree of substitution of Sn with Pb and/or In (Figure 1a). Due to the unchanged crystal structure, the crystal point group symmetries that are essential to maintain the topological surface states remain the same. Because of the difference in lattice constants of the end members (PbTe > SnTe  > InTe)\cite{Springer}, the lattice parameters of \psit\  compounds vary with $x$ and $y$. Figure 1b shows the XRD patterns of SnTe ($x=1$, $y=0$, black), \pstn\ ($x=0.5$, $y=0$, blue) and (Pb$_{0.5}$Sn$_{0.5}$)$_{0.7}$In$_{0.3}$Te ($x=0.5$, $y=0.3$, red), respectively. Compared to the parent compound SnTe, with a lattice constant $a=6.32$~\AA, Pb-doping increases the lattice constant ($a=6.39$~\AA\ for Pb$_{0.5}$Sn$_{0.5}$Te). Subsequent In-doping can then decrease the lattice constant ($a=6.36$~\AA\ for \psitn). Similarly,  a systematic shrinking of the unit cell as a function of In content has been observed in previous studies of SIT\cite{Zhong2013, Haldo2016}. The measured lattice parameters listed on the figure for the various compositions are qualitatively consistent with Vegard's law and the differences in radii of the ionic components. With more Sn or Pb atoms being replaced by In, the distortion of the crystal structure gets larger, and eventually the solubility limit is reached, indicated by the appearance of the secondary phase InTe.  

Common tools to characterize the surface states include angle-resolved photoemission spectroscopy (ARPES) and scanning tunneling microscopy (STM).  To apply these techiques, one typically needs atomically-flat and well-oriented surfaces. For the topological insulator Bi$_{2}$Se$_{3}$, this is not be a problem, since it can be easily cleaved due to weak coupling between its layers. In the case of SnTe related compounds, however, the situation is more challenging due to their isotropic cubic structures. To illustrate, Figures 1c-1e show microscope photos of pristine surfaces. The flat, shiny planes are cleaved surfaces, and they become smaller with increasing Pb/In substitution. Thus, it appears that substitution of Pb or In atoms introduces lattice distortion and leads to smaller cleaved surfaces for surface-sensitive studies. STM studies of SIT single crystal samples have been successfully performed\cite{Sato2013}, as discussed in Sec.~2.5.  Direct ARPES studies of PSIT single crystals are few, and it has proved more practical to perform measurements on thin films evaporated from previously characterized bulk samples\cite{Du2015, Zhong2015}.

\subsection{Resistivity behaviors of In-doped \pst}

\begin{figure}
\centering
\includegraphics[width=9cm]{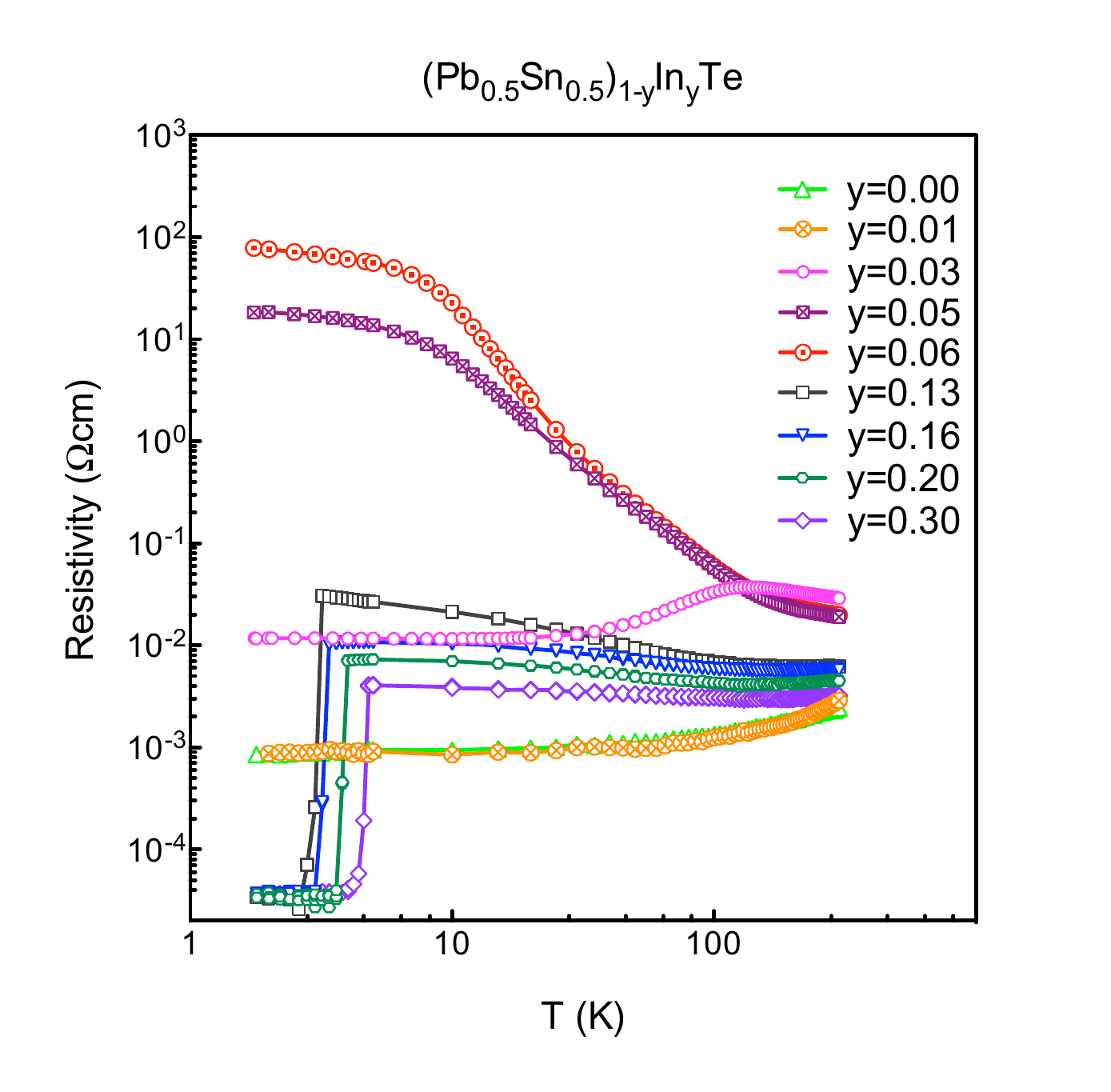}
\caption{\label{fig:R} (color online) Temperature dependence of the resistivity for (Pb$_{0.5}$Sn$_{0.5}$)$_{1-y}$In$_{y}$Te single crystals with indium contents $0\leq y \leq0.30$.}
\end{figure}   

The evolution of the electronic properties with composition have been investigated through transport measurements. Here we take (Pb$_{0.5}$Sn$_{0.5}$)$_{1-y}$In$_{y}$Te as an example to illustrate the effect of indium substitution. As shown in Figure 2, pure \pstn\ shows a metallic-like behavior with a $p$-type carrier density similar to SnTe. By introducing increasing amounts of indium into the \pstn\ system, single crystal samples show quite divergent, nonmonotonic variations in resistivity in the normal state. For the samples with one percent or less indium, the resistivity is weakly metallic, just like the resistivity behavior of pure SnTe\cite{Zhong2013} or \pst\ without indium doping\cite{Dixon1968}. Increasing $y$ to 0.06, we observe that the resistivity at 10~K rises by five orders of magnitude. With further increases of $y$, the resistivity drops, but remains semiconducting, consistent with earlier studies\cite{Zhong2014, Vul1978, Kozub2006, Shamshur2008}. This resistivity behavior in the normal state is quite different from the case of In doped SnTe\cite{Zhong2013}, where all samples are weakly metallic in the normal sate. At low temperature, samples show true bulk-insulating resistivity and and present evidence for nontrivial topological surface states\cite{Zhong2015}. With higher indium doping levels, superconductivity with a transition temperature \Tc\ positively correlated to the indium concentration was observed, and the highest \Tc, {$\sim4.7$~K, was achieved for 45\% indium doped SnTe samples \cite{Zhong2013,Balakrishnan2013} and 30\% indium doped Pb$_{0.5}$Sn$_{0.5}$Te samples \cite{Zhong2015}. 

\begin{figure}[!b]
\centering
\includegraphics[width=15cm]{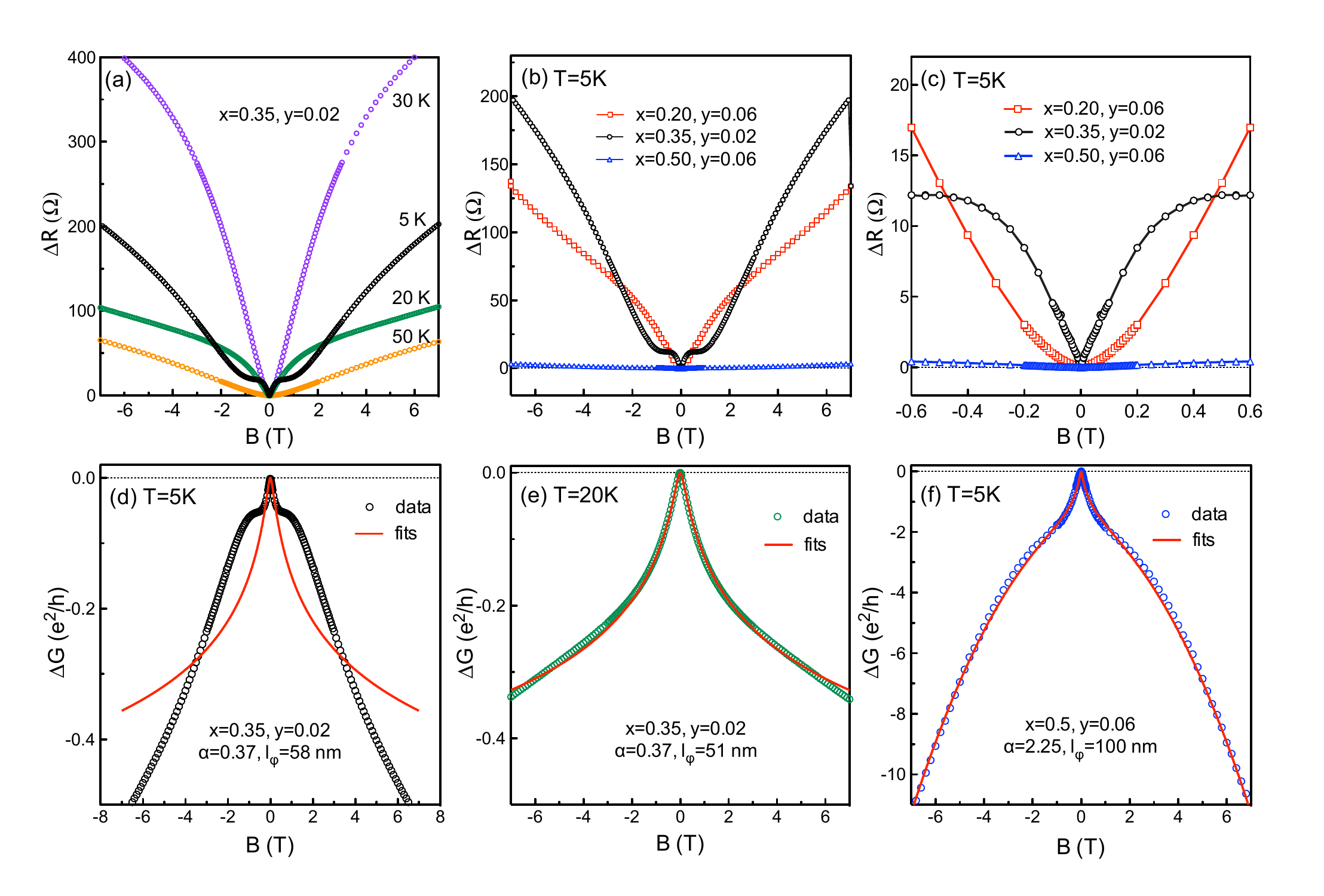}
\caption{\label{fig:RH} (color online) Weak antilocalization magnetoresistance of \psit.  (\textbf{a}) MR for a (Pb$_{0.65}$Sn$_{0.35}$)$_{0.98}$In$_{0.02}$Te sample measured at temperatures of $T= 5$, 20, 30 and 50 K in perpendicular magnetic fields of $|B|\leq 7$ T. The WAL effect is overwhelmed at high temperatures by the bulk conduction states. (\textbf{b,c}) MR for (Pb$_{0.65}$Sn$_{0.35}$)$_{0.98}$In$_{0.02}$Te (black), (Pb$_{0.5}$Sn$_{0.5}$)$_{0.94}$In$_{0.06}$Te (blue) and (Pb$_{0.8}$Sn$_{0.2}$)$_{0.94}$In$_{0.06}$Te (red) measured at 5 K in a full field range of $\pm7$ T and the enlarged low-field regime $|B|\leq 0.6$ T.  (\textbf{d,e}) Magnetoconductance $\Delta G=\Delta (1/R)$ of the (Pb$_{0.65}$Sn$_{0.35}$)$_{0.98}$In$_{0.02}$Te sample measured at 5~K and 20~K, respectively. Lines represent the result of fitting using WAL formula with fixed $\alpha=0.37$ and variable $l_\phi=51$~nm (20~K) and 58~nm (5~K). (\textbf{f}) Magnetoconductance of the (Pb$_{0.5}$Sn$_{0.5}$)$_{0.94}$In$_{0.06}$Te sample measured at 5~K. Line is fitted by the WAL formula together with an additional $-B^2$ term. }
\end{figure}   

The effect of indium substitution is similar for other (Pb,Sn)Te compositions. Nonmonotonic variation in the normal-state resistivity with $y$ is also found in transport measurements of PSIT for many series with different $x$ values. Specifically, $x$=0.5 is not the only system that shows large bulk resistance when doped with a low concentration of indium. In the whole family of \psit, maximum resistivities that surpass 10$^{6}\  \Omega$cm are observed for $x$=0.25-0.30. Even for $x$=0.35, doping with 6\% In results in a rise in resistivity by 6 orders of magnitude at low temperature. These phenomena can be well explained in a picture where the chemical potential is pinned within the band gap, which will be discussed in detail in a later section. 

A common test of the topological character of surface states involves measurements of magnetoresistance (MR) at low temperature\cite{Bansal2012}. The symmetry-protected coupling of spin and momentum for surface states makes them immune to weak localization effects.  Application of a transverse magnetic field breaks time-reversal symmetry\cite{Serbyn2014}, thus removing the topological protection and leading to a field-induced increase in resistance.  Figure 3a shows data for $\Delta R=R(B)-R(0)$ measured at several temperatures for a magnetic induction $|B|\leq 7$~T applied perpendicular to the $(001)$ surface of the (Pb$_{0.65}$Sn$_{0.35}$)$_{0.98}$In$_{0.02}$Te sample.  At temperatures of 30~K and below, the field dependence of the induced resistance has a form qualitatively consistent with that expected for weak anti-localization (WAL) of two-dimensional electron states.The MR curve at 5K (black) clearly shows a cusp near zero field, which is a sign of the WAL effect and suggests the dominance of topological surface states. At elevated temperature, the cusp disappears, and the curves in the low-field regime (not shown) are dominated by the parabolic $B$-dependence of the bulk states\cite{Akiyama2014, Kim2011WAL}, which is a reflection of the bulk carriers under a Lorentz force in a perpendicular field. The magnitude of the MR changes monotonically with temperature, a fact that needs further study to fully understand. 

In order to clarify the nature of surface states in samples with different compositions, in Fig. 3b, 3c we compare the MR behavior at 5 K between (Pb$_{0.8}$Sn$_{0.2}$)$_{0.94}$In$_{0.06}$Te (red), (Pb$_{0.65}$Sn$_{0.35}$)$_{0.98}$In$_{0.02}$Te (black), and (Pb$_{0.5}$Sn$_{0.5}$)$_{0.94}$In$_{0.06}$Te (blue). In the low-field regime, the $x$=0.35 sample clearly shows a WAL effect even with a few percent indium, which is consistent with the ARPES evidence that the topological surface states of In-doped SnTe are maintained when In doping concentration is roughly 4.5\%\cite{Sato2013}. To be more quantitative, we convert the data to conductance, $G$, and compare with the theoretical formula for WAL\cite{Hikami1980},
\begin{equation}
  \Delta G = {\alpha\over\pi} {e^2\over h} [\ln(B_\phi/B) - \psi({\textstyle\frac12}+B_\phi/B)],
\end{equation}
where $\psi$ is the digamma function and $\alpha$ is a number equal to $1/2$ times the number of conduction channels; $B_\phi= \Phi_0/(8\pi l_\phi^2)$, with $\Phi_0=h/e$ and $l_\phi$ being the electronic phase coherence length.   For our system, one expects four Dirac cones crossing the Fermi surface\cite{Serbyn2014, Assaf2014}, which would give $\alpha=2$.  Figure 3b shows that we get a good fit to the 20-K data for the $x=0.35$, $y=0.02$ sample with $\alpha=0.37$ and $l_\phi=51$~nm.  Moving to $T=5$~K in Fig.~3d, the low-field data can be described by keeping $\alpha$ fixed and increasing $l_\phi$ to 58 nm; however, the data also exhibit a large oscillation about the calculated curve for $|B|>0.2$~T.  This may be due to a Landau level crossing the Fermi energy\cite{Serbyn2014}.  Turning to the $x=0.5$, $y=0.06$ sample, the 5-K data in Fig.~3d are well described by the WAL formula for $|B|<1$~T, with $\alpha=2.25$ and $l_\phi=100$~nm, but at larger $|B|$ we need an additional component that varies as $-B^2$.  The latter contribution might come from bulk states.  

\subsection{Phase diagram}

To summarize the effect of indium substitution on PSIT materials, we present in Fig.~4, a ternary phase diagram of the system to illustrate trends for several properties: the character of the low-temperature resistivity (metallic, insulating, superconducting) and the solubility limit. Here, the end members are SnTe, PbTe, and InTe.  The closer to the end member, the higher concentration of that component. Each of the six dashed lines starting with the same end member InTe represent a series of PSIT with the fixed Sn:Pb ratios, as labeled by $x$.  

For low indium doping (blue region), samples show weak metallic resistivity, as in SnTe. A few percent indium doping turns the \pst\ samples into true insulators (orange region). By increasing the In content further, superconductivity may be achieved (green region). When the indium content exceeds the solubility limit in the system (marked with white crosses), where additional In is no longer simply substituting for Pb/Sn, an impurity phase of InTe, with a tetragonal crystal structure, appears and the samples are no longer single crystals. The critical In concentrations that divide these various regions are illustrated with dashed lines. 

\begin{figure}
\centering
\includegraphics[width=10cm]{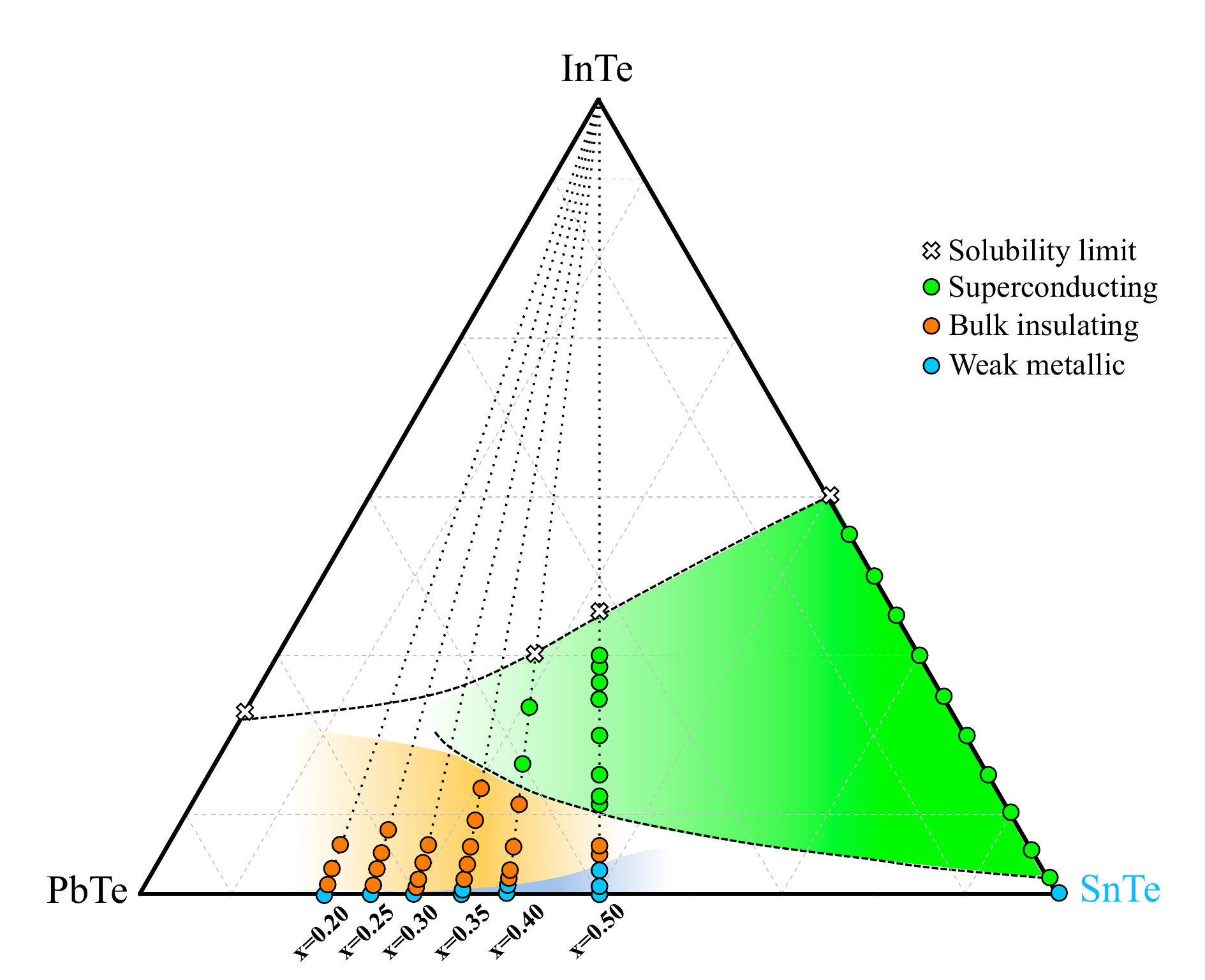}
\caption{\label{fig:diagram} (color online) A ternary phase diagram summarizing all the resistivity behaviors of \psit. Experimental results for SIT with In content up to 10\% is obtained from Ref. \cite{Erickson2009}. The solubility limit of In in PbTe (24\%) is obtained from Ref. \cite{Ravich2002}. Samples with weak metallic resistivity are shown in blue, with insulating resistivity are shown in orange, and with superconductivity are shown in green. White crosses represent the solubility limit of In, beyond which the sample no longer remains in a single phase and secondary InTe phase shows up.}
\end{figure}   

From the resistivity behavior in the phase diagram, it can be seen that the In substitution effect shows consistent trends. Superconductivity emerges almost immediately with indium doping in SnTe. In \pst, though, with increasing Pb content the amount of In needed to induce superconductivity goes up, and the range of superconductivity with respect to In-doping shrinks. Meanwhile, the bulk insulating region is broadens with increased Pb, and the maximum bulk resisitvity that can be achieved in the PSIT family is found in the $x$=0.30 and $x$=0.25 series\cite{Zhong2015}. Those materials along with previous reported bulk insulating TIs Sn-Bi$_{1.1}$Sb$_{0.9}$Te$_{2}$S\cite{Kushwaha2016} could provide good platforms to study the true topological 'insulators', in which bulk conduction would not dominate the transport behavior, assuming that their surface states remain topological.

\subsection{Bulk band structure}

\begin{figure}[!t]
\centering
\includegraphics[width=12cm]{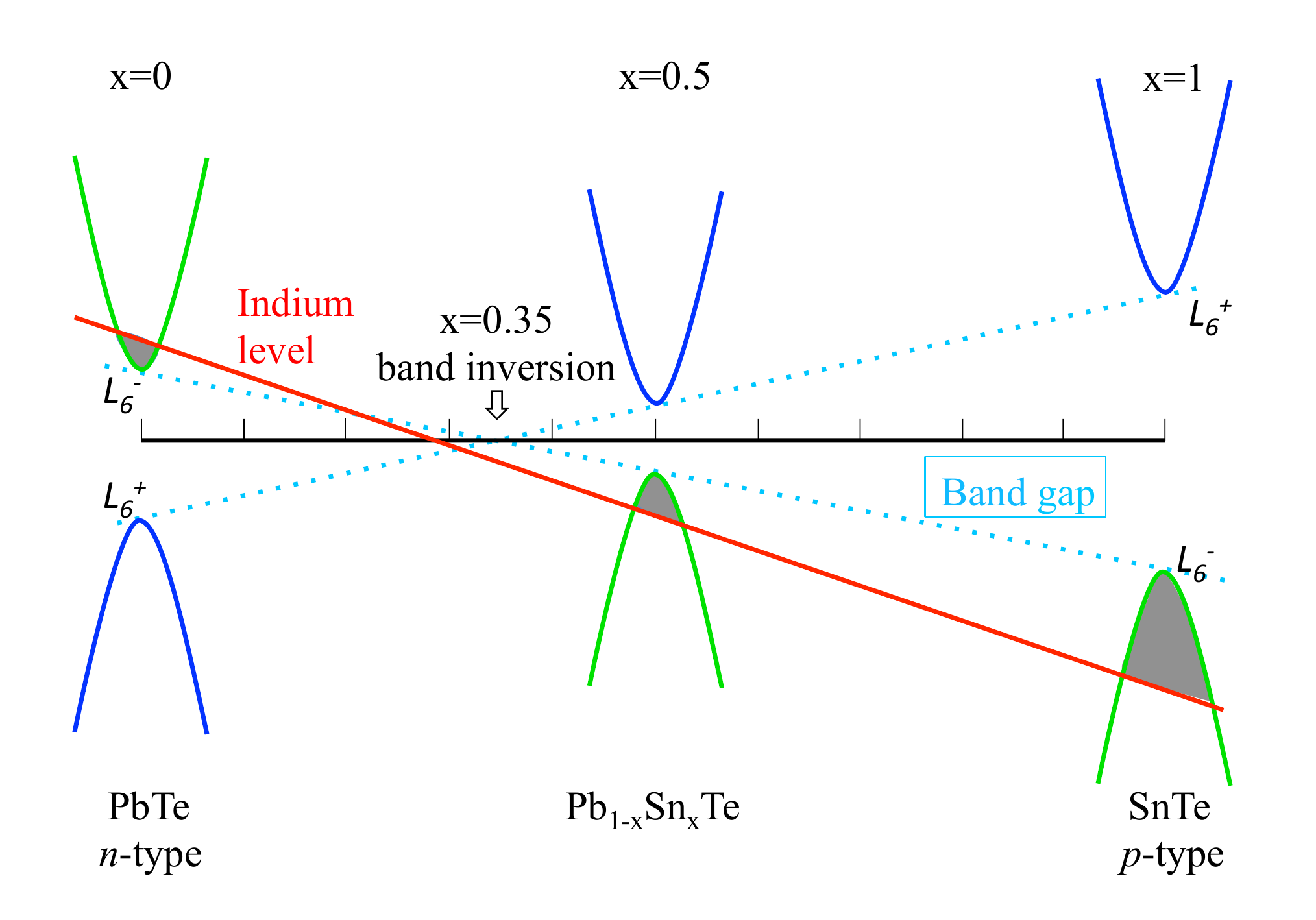}
\caption{\label{fig:Band} (color online) Energy diagrams illustrating the relative location of the conduction, valence band and indium induced impurity band in the continuous series of \pst\ alloys with low In doping level, where indium can be simply treated as a $p$-type dopant. In SnTe, the conduction band has a symmetry of $L_{6}^{+}$; this undergoes a band inversion at $x\sim$0.35 and the symmetry is inverted in PbTe. The band gap is illustrated with blue dashed lines, with the end member SnTe having 360 meV and PbTe having 190 meV \cite{Heremans2012}.  The Fermi level, controlled by the indium impurity states, is indicated schematically by the red line. }
\end{figure}   

To address the divergent resistivity behaviors, it is helpful to consider the bulk electronic structure.  SnTe and other IV-VI materials with the rocksalt structure have long attracted attention as a model for small band gap semiconductors.  The topologically distinct band structure of SnTe (nontrivial, $x$ = 1) and PbTe (trivial, $x$ = 0) involves a change in the ordering of the conduction and valence bands at $L$ points.  This implies that the band gap of the alloy \pst\ first closes and then re-opens as $x$ increases, as shown in Fig.~5\cite{Ravich2002, Kaidanov1985, Pankratov1987}.  It follows that there must be a topological quantum phase transition upon varying the Pb/Sn ratio in \pst, and experiments indicate that it occurs near $x_{c}\approx 0.35$ at low temperature\cite{Dimmock1966, Kaidanov1985, Pankratov1987, Gao2008, Xu2012, Tanaka2013}. 

Generally, it is believed that each In dopant will provide one less valence electron than Sn$^{2+}$ and Pb$^{2+}$, so that indium should be considered as a $p$-type dopant.  In the case of SnTe, one begins with a $p$-type semiconductor due to Sn vacancies.  With In doping, the number of cation vacancies decreases, which partially compensates the expected impact of the In; nevertheless, the $p$-type carrier density initially grows with increasing In concentration \cite{Zhang2013_SnTe,Novak2013}.  The situation becomes more complicates for an Indium concentration above 10\%, where the sign of the Hall resistivity changes \cite{Haldo2016}, suggesting the possibility that two types of carriers are simultaneously present.  In fact, in indium doped PbTe and \pst, In-doping results in far less than 1 electron per impurity atom, which suggests In doping also introduces an impurity band that causes the effect of pinning the Fermi level\cite{Bushmarina1991, Heremans2012}. Evidence for the quasi-localized character of indium-induced states has been provided by a recent nuclear magnetic resonance study on Sn$_{0.9}$In$_{0.1}$Te\cite{Maeda2017}. 

In this scenario, the large bulk resistivity in series with $x=0.25-0.35$ is a consequence of indium sites introducing localized impurity states that pin the chemical potential\cite{Kaidanov1985, Ravich2002}; the electronic properties then depend on the position and width of the indium level. In the region of compositions where the localized impurity band lies in the band gap, or a position that is very close to the band edge, the free carrier concentration is extremely low at low temperature, which is reflected in the very large bulk resistivities for $x\sim0.30$ that we observe in the transport measurements. 

\begin{figure}[!t]
\centering
\includegraphics[width=15cm]{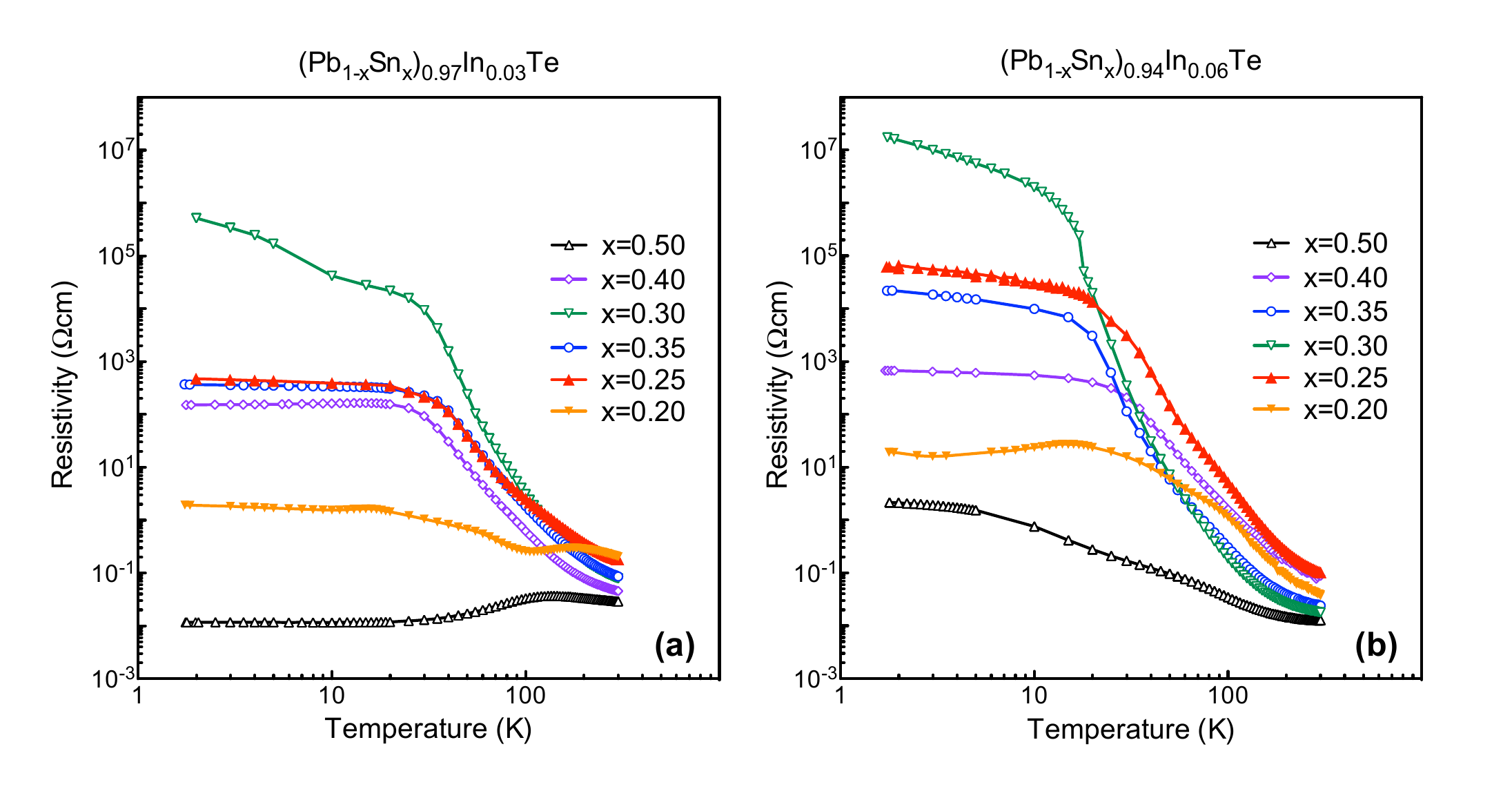}
\caption{\label{fig:R2} (color online) Temperature dependence of the resistivity for (Pb$_{1-x}$Sn$_{x}$)$_{0.97}$In$_{0.03}$Te (a) and (Pb$_{1-x}$Sn$_{x}$)$_{0.94}$In$_{0.06}$Te (b) single crystals. The resistivity value are shown in a logarithmic scale. }
\end{figure}   

According to the schematic evolution of the band structure of \pst\ and the energy of the In impurity level in Fig.~5,  the chemical potential sits in the valence band on the Sn side, consistent with $p$-type metallic behavior, while it moves to the conduction band on the Pb-rich side. With a very small amount of indium doping on the Pb-rich side, the opposing trends of decreasing cation vacancies and increasing In-substitution initially lower the carrier density, leaving the system weakly metallic. With further increases in In content, the Fermi level drops into the band gap, where it gets pinned by the impurity state level.  The magnitude of the resistivity will then depend largely on the size of the band gap, which is determined by the Sn content, $x$, instead of the indium content, $y$\cite{Ravich2002}.  Figure 6 gives a summary of the variation in resistivity as a function of $x$ in 
PSIT compounds with either 3\% or 6\% indium doping.  The same trends are found as a function of $x$, although the low-temperature resistivities tend to be higher for $y=0.06$. 

It is worth mentioning that a long relaxation time was observed in the bulk resistance for several samples, especially those that are truly bulk insulators, i.e. $x=0.25$, 0.30, 0.35. After a sample was quenched down to low temperature (liquid-helium), its resistivity gradually decreased with time. This relaxation phenomenon can last for days until the resistivity reaches a stable value. Previous studies on \pst\ doped with group-III elements revealed similar time-dependent behavior, and it was explained in terms of the interaction between the crystal lattice and the non-equilibrium electron densities associated with the pinned chemical potential at the impurity level\cite{Kaidanov1985}. 

\subsection{Debate on topological superconductivity}
At higher indium content (>10\%), superconductivity emerges in \pst\ samples, with a typical superconducting transition temperature in the range of 3 to 5~K. There are intriguing questions about the nature of the superconductivity: is it conventional BCS superconductivity, or unconventional topological superconductivity? Topological superconductors are accompanied by gapless states at the edge or surface, which characterize the nontrivial topology of the bulk state and they may be composed of Majorana fermions 
\cite{Qi2009, Qi2010}. 

The first plausible example of TSC (associated with TI or TCI compounds) was Cu$_{x}$Bi$_{2}$Se$_{3}$\cite{Fu2010odd}. Experimental evidence from point-contact spectroscopy\cite{Sasaki2011, Kirzhner2012} showing zero-bias conductance peaks coexisting with a superconducting gap may be indicative of the unconventional superconductivity, which is necessary (but not sufficient) for TSC in inversion symmetric, time-reversal-invariant superconductors. Similarly, results for In-doped SnTe from both point-contact spectroscopy and high-resolution ARPES studies have been interpreted as evidence   for odd-parity pairing and topological superconductivity \cite{Sasaki2012, Sato2013}. 

A markedly different conclusion was drawn, however, in an STM study on Cu$_{0.2}$Bi$_{2}$Se$_{3}$\cite{Levy2013}, which reported a superconducting gap without any zero bias anomalies.  Later studies on the optimally doped TCI system SIT using thermal conductivity\cite{He2013}, magnetization and muon-spin rotation ($\mu$SR) measurements\cite{Saghir2014} also supported the conclusion that SIT has a ful superconducting gap in the bulk, and is more likely to be a conventional $s$-wave superconductor. Similarly, STM measurements\cite{Du2015} of the superconducting state as well as the superconducting energy gap in \psitn\ on the high-symmetry (001) surface lead to the same conclusion, that the superconducting sample seems to be fully gapped without any in-gap states, contrary to the expectations for a topological superconductor. 

These controversies may be due to the complexity of the junctions in point contact measurements, since the spectra that are indicative of an unconventional superconductor can also be interpreted by other mechanisms\cite{Kirzhner2012, Du2015}. On the other hand, the observed fully-gapped tunneling spectra in STM measurements on Cu$_{x}$Bi$_{2}$Se$_{3}$ and SIT can be also explained by the results of exotic pairing states with additional parameters\cite{Levy2013}. In addition, in TCI compounds  where the exotic surface states only exist on certain high-symmetry planes guaranteed by the mirror symmetry, the possibility of topological superconductivity feature cannot be ruled out from  studies of the (001) plane alone \cite{Du2015}. Besides, due to the poor cleavability cubic \psit, it might be tricky to expose the desired surface for surface-sensitive measurements.

The debate on topological superconductivity has recently been reinvigorated by a nuclear magnetic resonance study of Cu$_{0.3}$Bi$_{2}$Se$_{3}$ \cite{Matano2016}. There the authors find clear evidence for a breaking of the spin-rotation symmetry in the superconducting state, consistent with spin-triplet pairing. This will surely motivated further investigations. 

\section{Discussion}
In recent years, In-doped SnTe and \pst\ have been studied extensively both as examples of topological crystalline insulators and potentially-interesting superconductors. Our study of a broad range of compositions in the PSIT system shows that indium doping has a nonmonotonic effect on the electronic properties, which can be explained from the standpoint of the relative location of the indium-induced impurity band and the bulk band structure. In this article we have presented a summary which recaps our findings and conclusions, which can be instructive for future work on this system. 

In the effort of looking for a new topological superconductor, Tl$_{5}$Te$_{3}$ has been found to be tunable between superconducting and topological surface states by Sn-substitution\cite{Arpino2014, Arpino2015}, which is quite similar to the In-substitution effect on the \pst\ system. These facts may imply that the topological surface states and the bulk superconductivity are two competing parameters. The goal of mixing the superconducting and topological characters remains a challenge. 

A plausible strategy of looking for Majorana fermions is to artificially construct topological insulator/conventional superconductor heterostructures and make of use the superconducting proximity effect\cite{Fu2008, Wang2012, Wang2013, Xu2014, Li2014, Xu2015}. Both SIT and PSIT would be perfect platforms for this purpose, since these systems undergo a continuous change from a TCI to a (likely conventional) superconductor. More specifically, the large bulk resistivity shows up in \psit\ with $x=0.25$--0.5, a $p$-type matrix can be realized in series the with $x=0.35$--1.0, and superconductivity can also be realized in the latter compounds, which makes this system quite promising for exploitation in heterostructures. 

\section{Experimental Section}

\subsection{Sample preparation.}

In the Pb-Sn-In-Te alloy system, several elemental metals or compounds are found to be superconducting near liquid-helium temperatures, as indicated in Fig. 7. To investigate the indium substitution effect on the (Pb,Sn)Te system, compounds with the composition in the area of the red triangle have been grown and carefully studied in this work. Single crystal samples with nominal composition, \psit\ ($x_{norm}$=0.2-0.5, 1.0, $y_{norm}$=0-0.5), were prepared via the vertical Bridgman method. Stoichiometric mixtures of high-purity (99.999\%) elements were sealed in double-walled evacuated quartz ampoules. The ampoules were set in a vertical position, heated at 950\degrees\ in a three-zone vertical box furnace, with rocking to achieve homogeneous mixing of the ingredients. The crystal growth took place via slow cooling from 950 to 760 \degrees\ at the rate of 1.5 \degrees /hr, and then the samples were gradually cooled down to room temperature over another 3 days\cite{Zhong2013}. Figure 8a shows examples of the resulting crystals (before removal from the quartz tubes). 

For a few compositions, where we needed large crystals for ARPES measurements, we used the modified floating-zone method. In the normal traveling-solvent floating-zone (TSFZ) method, a polycrystalline rod is prepared as feed rod and seed rod, fixed at the upper or lower shaft of the furnace. During the growth, the molten zone is not in contact with any container. However, elements such as indium and tin easily evaporate and are reactive when heated to high temperature. Therefore, we used the modified TSFZ method, as explained in detail in our earlier publication\cite{Zhong2014}, where the starting material, prepared in the vertical Bridgman method, was sealed in a long quartz tube ($\sim$15 cm) and mounted at the bottom shaft in the floating zone furnace. The space inside the large glass chamber surrounding the quartz tube was filled with high-purity Ar at 1 bar to avoid oxygen diffusion through the quartz. During the growth the rotating shaft kept moving downwards at a velocity of 0.5-1 $mm/hr$, so that the new crystal gradually grew from the bottom of the starting ingot, resulting in a sample such as that shown in Fig.~8b. 

\begin{figure}
\centering
\includegraphics[width=10cm]{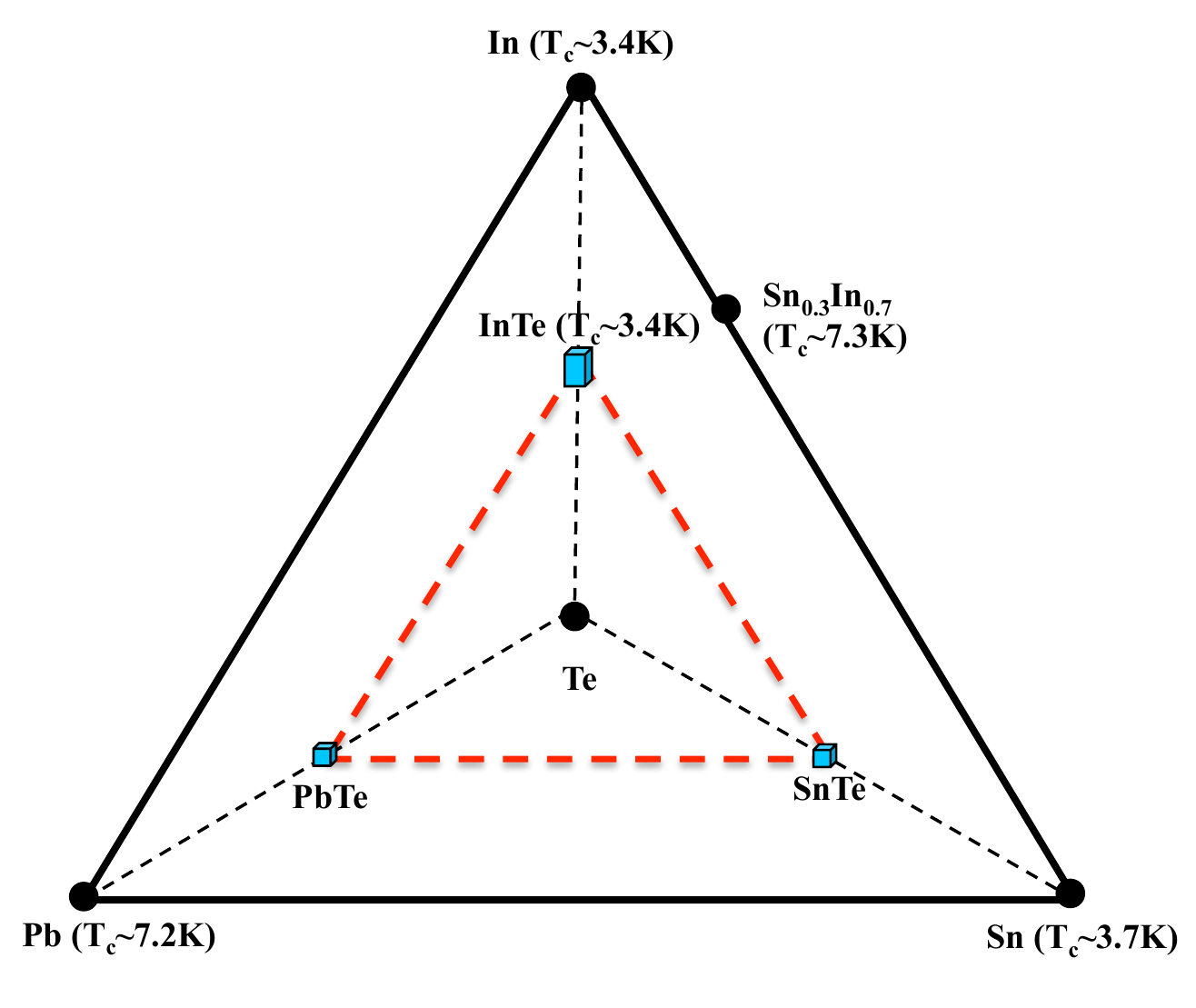}
\caption{\label{fig:Phase2} (color online) Phase diagram of composition and superconductivity of the Pb-Sn-In-Te alloy system. The known superconducting metals or compounds Pb, Sn, In, InTe, and Sn$_{0.3}$In$_{0.7}$ are marked on the diagram, respectively. Based on this superconducting phase diagram, we did thorough studies in the region marked with red dashed lines. }
\end{figure}   

For each sample used in the magnetization, transport and other measurements, the chemical composition was characterized using energy-dispersive X-ray spectroscopy (EDS). In this article, chemical composition values $x$ and $y$ correspond to the measured concentrations. 

\begin{figure}
\centering
\includegraphics[width=15cm]{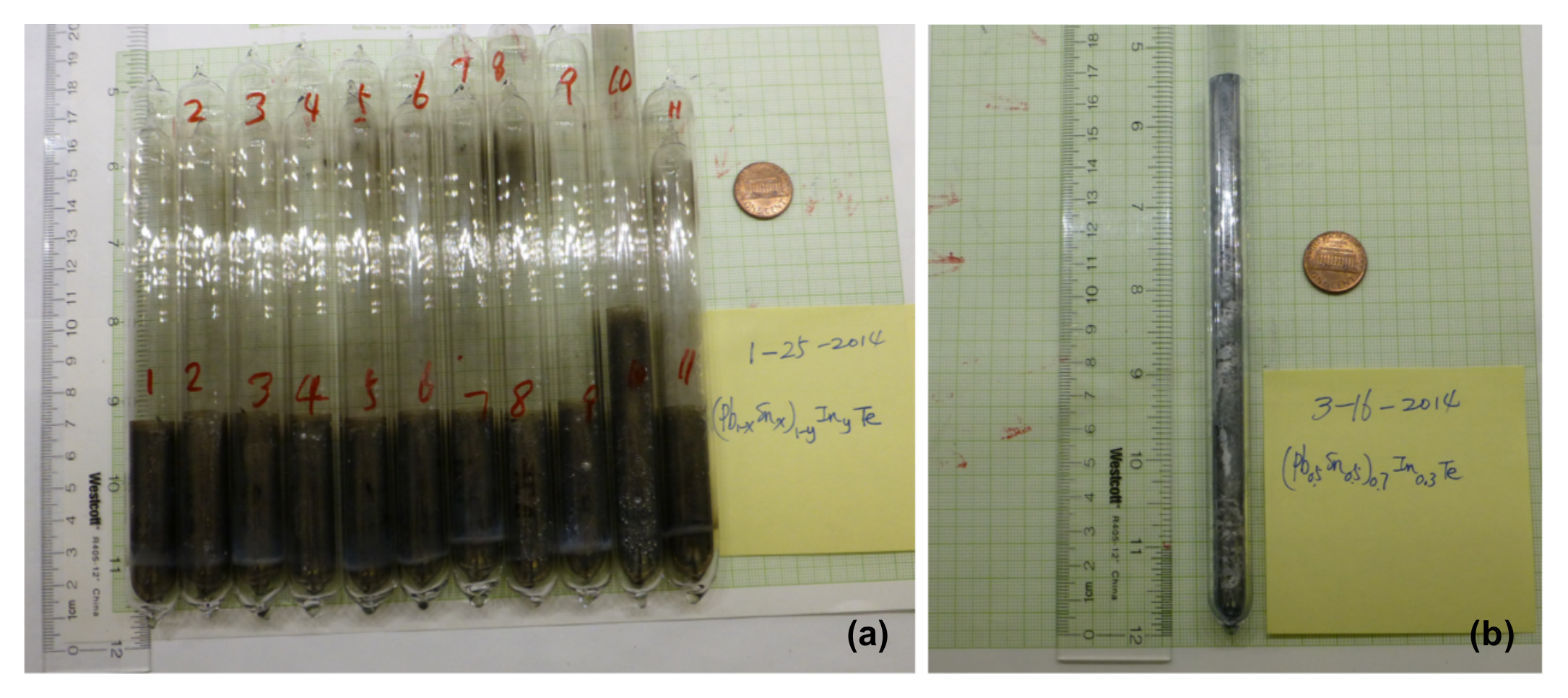}
\caption{\label{fig:Ingot} (color online) Single crystal rods of PSIT alloy grown by vertical Bridgman method (\textbf{a}) and modified floating zone method (\textbf{b}).}
\end{figure}   

\subsection{Sample characterizations.}

To identify the room-temperature crystal structure, each sample was characterized by X-ray powder diffraction (XRD) measured with Cu $K \alpha$ radiation from a model Rigaku Miniflex II. Microstructure and chemical composition of the samples were carefully investigated by an analytical high-resolution scanning electron microscope (SEM) equipped for EDS, model JEOL 7600F, located at the Center for Functional Nanomaterials (CFN) at Brookhaven National Laboratory. For each crystal piece, EDS was measured at 10 positions and the mean value was taken to characterize the sample. These measured $x$ and $y$ values agreed within the measurement uncertainty ($\pm$0.02), and the measured values are used throughout this article. Typical microstructure pictures of the PSIT cleaved surfaces were taken by optical microscope.

To study the effect of indium substitution on the magnetic properties, dc magnetic susceptibility measurements were performed using a commercial superconducting quantum interference device (SQUID) magnetometer (MPMS, Quantum Design), for temperatures down to 1.75 K. The sample pieces were cut into an approximately cubic shape, typically weighing 0.1 g. 

For transport measurements, thin bar-like samples with typical dimensions of $4 \times1.5\times0.5$ mm$^3$ were cut from the bulk crystal and then polished. Electrical resistance was measured in the standard four-probe configuration, using gold wires and room-temperature-cured, fast-drying silver paint for the ohmic contacts on top side, and performed with a Keithley digital multimeter (model 2001), using the MPMS for temperature control. Measurement errors due to the contact geometry are estimated to be less than 10\%. 

\begin{acknowledgments}
Work at Brookhaven National Laboratory is supported by the U.S. Department of Energy, Office of Basic Energy Sciences, Division of Materials Sciences and Engineering, under Contract No. DE-SC0012704; use of facilities at the Center for Functional Nanomaterials was supported by the Office of Basic Energy Sciences, Division of Scientific User Facilities. W. K. acknowledges support from National Natural Science Foundation of China \#11674220 and 11447601, and Ministry of Science and Technology  \#2016YFA0300500 and 2016YFA0300501..
\end{acknowledgments}


\renewcommand\bibname{References}


\begin{thebibliography}{999}
\bibitem{Hasan2010}
Hasan, M. Z. and Kane, C. L.. Colloquium: Topological insulators. {\em Rev. Mod. Phys.} {\bf 2010}, {\em 82}, 3045.

\bibitem{Qi2011}
Qi, X.-L. and Zhang, S.-C.. Topological insulators and superconductors. {\em Rev. Mod. Phys.} {\bf 2011}, {\em 83}, 1057.

\bibitem{Ando2013}
Ando, Yoichi. Topological insulator materials. {\em J. Phys. Soc. Jpn.} {\bf 2013}, {\em 82}, 102001.

\bibitem{Kane2011}
Kane, Charles and Moore, Joel. Topological insulators. {\em Physics World} {\bf 2011}, {\em 24}, 32.

\bibitem{Fu2007}
Fu, L.; Kane, C. L.; and Mele, E. J.. Topological insulators in three dimensions. {\em Phys. Rev. Lett.} {\bf 2007}, {\em 98}, 106803.

\bibitem{Bernevig2006}
Bernevig, B. A.; Hughes, T. L.; and Zhang, S. C.. Quantum spin Hall effect and topological phase transition in HgTe quantum wells. {\em Science} {\bf 2006}, {\em 314}, 1757.

\bibitem{Hsieh2008}
Hsieh, D.; Qian, D.; Wray, L.; Xia, Y.; Hor, S.; Cava, R. J.; and Hasan, M. Z.. A topological Dirac insulator in a quantum spin Hall phase. {\em Nature} {\bf 2008}, {\em 454}, 970.

\bibitem{Zhang2009}
Zhang, H.; Liu, C.-X.; Qi, X.-L.; Dai, X.; Fang, Z.; and Zhang, S.-C.. Topological insulators in Bi$_{2}$Se$_{3}$, Bi$_{2}$Te$_{3}$ and Sb$_{2}$Te$_{3}$ with a single Dirac cone on the surface. {\em Nat. Phys.} {\bf 2009}, {\em 5}, 438.

\bibitem{Xia2009}
Xia, Y.; Qian, D.; Hsieh, D.; Wray, L.; Pal, A.; Bansil, A.; Grauer, D.; Hor, Y. S.; Cava, R. J.; and Hasan, M. Z.. Observation of a large-gap topological-insulator class with a single Dirac cone on the surface. {\em Nat. Phys.} {\bf 2009}, {\em 5}, 398.

\bibitem{Chen2009}
Chen, Y. L.; Analytis, J. G.; Chu, J.-H.; Liu, Z. K.; Mo, S.-K.; Qi, X. L.; Zhang, H. J.; Lu, D. H.; Dai, X.; Fang, Z.; Zhang, S. C.; Fisher, I. R.; Hussain, Z.; and Shen, Z.-X.. Experimental realization of a three-dimensional topological insulator, Bi$_{2}$Te$_{3}$. {\em Science} {\bf 2009}, {\em 325}, 178.

\bibitem{Fu2011}
Fu, L.. Topological crystalline insulators. {\em Phys. Rev. Lett.} {\bf 2011}, {\em 106}, 106802.

\bibitem{Hsieh2012}
Hsieh, Timothy H.; Lin, Hsin; Liu, Junwei; Duan, Wenhui; Bansil, Arun; and Fu, Liang. Topological crystalline insulators in the SnTe material class. {\em Nat. Commun.} {\bf 2012}, {\em 3}, 982.

\bibitem{Tanaka2012}
Tanaka, Y.; Ren, Zhi; Sato, T.; Nakayama, K.; Souma, S.; Takahashi, T.; Segawa, Kouji; and Ando, Yoichi. Experimental realization of a topological crystalline insulator in SnTe. {\em Nat. Phys.} {\bf 2012}, {\em 8}, 800.

\bibitem{Dziawa2012}
Dziawa, P.; Kowalski, B. J.; Dybko, K.; Buczko, R.; Szczerbakow, A.; Szot, M.; \L{}usakowska, E.; Balasubramanian, T.; Wojek, B. M.; Berntsen, M. H.; Tjernberg, O.; and Story, T.. Topological crystalline insulator states in Pb$_{x}$Sn$_{x}$Se. {\em Nat. Mater.} {\bf 2012}, {\em 11}, 1023.

\bibitem{Xu2012}
Xu, Su-Yang; Liu, Chang; Alidoust, N.; Neupane, M.; Qian, D.; Belopolski, I.; Denlinger, J.D.; Wang, Y. J.; Lin, H.; Wray, L.A.; Landolt, G.; Slomski, B.; Dil, J. H.; Marcinkova, A.; Morosan, E.; Gibson, Q.; Sankar, R.; Chou, F. C.; Cava, R. J.; Bansil, A.; and Hasan, M. Z.. Observation of a topological crystalline insulator phase and topological pahse transition in \pst. {\em Nat. Commun.} {\bf 2012}, {\em 3}, 1192.

\bibitem{Yan2014}
Yan, Chenhui; Liu, Junwei; Zang, Yunyi; Wang, Jianfeng; Wang, Zhenyu; Wang, Peng; Zhang, Zhi-dong; Wang, Lili; Ma, Xucun; Ji, Shuaihua, He, Ke; Fu, Liang; Duan, Wenhui; Xue, Qi-kun; and Chen, Xi. Experimental observation of Dirac-like surface states and topological phase transition in \pst (111) films. {\em Phys. Rev. Lett.} {\bf 2014}, {\em 112}, 186801.

\bibitem{Sun2013}
Sun, Yan; Zhong, Zhicheng; Shirakawa, Tomonori; Franchini, Cesare; Li, Dianzhong; Li, Yiyi; Yunoki, Seiji; and Chen, Xing-Qiu. Rocksalt SnS and SnSe: Native topological crystalline insulators. {\em Phys. Rev. B} {\bf 2013}, {\em 88}, 235122.

\bibitem{Barone2013}
Barone, Paolo; Rauch, Tom\'{a}\v{s}; Sante, Domenico Di; Henk, J\"{u}rgen; Mertig, Ingrid; and Picozzi, Silvia. Pressure-induced topological phase transitions in rocksalt chalcogenides. {\em Phys. Rev. B} {\bf 2013}, {\em 88}, 045297.

\bibitem{Hsieh2014}
Hsieh, Timothy H.; Liu, Junwei; and Fu, Liang. Topological crystalline insulators and Dirac octets in antiperovskites. {\em Phys. Rev. B} {\bf 2014}, {\em 90}, 081112(R).

\bibitem{Munoz2016}
Munoz, F.; Vergniory, M. G.; Rauch, T.; henk, J.; Chulkov, E. V.; Mertig, I.; Botti, S.; Marques, M. A. L.; and Romero, A. H.. Topological crystalline insulator in a new Bi semiconducting phase. {\em Sci. Rep.} {\bf 2016}, {\em 6}, 21790.

\bibitem{Fu2010odd}
Fu, Liang and Berg, Erez. Odd-parity topological superconductors: Theory and application to Cu$_{x}$Bi$_{2}$Se$_{3}$. {\em Phys. Rev. Lett.} {\bf 2010}, {\em 105}, 097001.

\bibitem{Hor2010sc}
Hor, Y. S.; Williams, A. J.; Checkelsky, J. G.; Roushan, P.; Seo, J.; Xu, Q.; Zandbergen, H. W.; Yazdani, A.; Ong, N. P.; and Cava, R. J.. Superconductivity in Cu$_{x}$Bi$_{2}$Se$_{3}$ and its implications for pairing in the undoped topological insulator. {\em Phys. Rev. Lett.} {\bf 2010}, {\em 104}, 057001.

\bibitem{Sato2013}
Sato, T.; Tanaka, Y.; Nakayama, K.; Souma, S.; Takahashi, T.; Sasaki, S.; Ren, Z.; Taskin, A. A.; Segawa, Kouji; and Ando, Yoichi. Fermiology of the strongly spin-orbit coupled superconductor \sit: Implications for topological superconductivity. {\em Phys. Rev. Lett.} {\bf 2013}, {\em 110}, 206804.

\bibitem{Erickson2009}
Erickson, A. S.; Chu, J.-H.; Toney, M. F.; Geballe, T. H.; and Fisher, I. R.. Enhanced superconducting pairing interaction in indium-doped tin telluride. {\em Phys. Rev. B} {\bf 2009}, {\em 79}, 024520.

\bibitem{Balakrishnan2013}
Balakrishnan, G.; Bawden, L.; Cavendish, S. and Lees, M. R.. Superconducting properties of the In-substituted topological crystalline insulator SnTe. {\em Phys. Rev. B} {\bf 2013}, {\em 87}, 140507(R).

\bibitem{Zhong2013}
Zhong, R. D.; Schneeloch, J. A.; Shi, X. Y.; Xu, Z. J.; Zhang, C.; Tranquada, J. M.; Li, Q.; and Gu, G. D.. Optimizing the superconducting transition temperature and upper critical field of \sit. {\em Phys. Rev. B} {\bf 2013}, {\em 88}, 020505(R).

\bibitem{Novak2013}
Novak, Mario; Sasaki, Satoshi; Kriener, Markus; Segawa, Kouji; and Ando, Yoichi. Unusual nature of fully gapped superconductivity in In-doped SnTe. {\em Phys. Rev. B} {\bf 2013}, {\em 88}, 140502(R).

\bibitem{Zhong2014}
Zhong, R. D.; Schneeloch, J. A.; Liu, T. S.; Camino, F. E.; Tranquada, J. M.; and Gu, G. D.. Superconductivity induced by In substitution into the topological crystalline insulator \pstn. {\em Phys. Rev. B} {\bf 2014}, {\em 90}, 020505(R).

\bibitem{Skinner2012}
Skinner, Brian; Chen, Tianran; and Shklovskii, B.I.. Why is the bulk resistivity of topological insulators so small? {\em Phys. Rev. Lett.} {\bf 2012}, {\em 109}, 176801.

\bibitem{Butch2010}
Butch, N.P.; Kirshenbaum, K.; Syers, P.; Sushkov, A.B.; Jenkins, G.S.; Drew, H.D.; and Paglione, J.. Strong surface scattering in ultrahigh-mobility Bi$_{2}$Se$_{3}$. {\em Phys. Rev. B} {\bf 2010}, {\em 81}, 241301(R).

\bibitem{Jia2011}
Jia, Shuang; Ji, Huiwen; Climent-Pascual, E.; Fuccillo, M.K.; Charles, M.E.; Xiong, Jun; Ong, N.P.; and Cava, R.J.. Low-carrier-concentration crystals of the topological insulator Bi$_{2}$Te$_{2}$Se. {\em Phys. Rev. B} {\bf 2011}, {\em 84}, 235206.

\bibitem{Peng2009}
Peng, Hailin; Lai, Keji; Kong, Desheng; Meister, Stefan; Chen, Yulin; Qi, Xiao-Liang; Zhang, Shou-Cheng; Shen, Zhi-Xun; and Cui, Yi. Aharonov-Bohm interference in topological insulator nanoribbons. {\em Nat. Mater.} {\bf 2011}, {\em 9}, 225.

\bibitem{Hor2009}
Hor, Y.S.; Richardella, A.; Roushan, P.; Xia, Y.; Checkelsky, J.G.; Yazdani, A.; Hasan, M.Z.; Ong, N.P.; and Cava, R.J.. $p$-type Bi$_{2}$Se$_{3}$ for topological insulator and low-temperature thermoelectric applicaitons. {\em Phys. Rev. B} {\bf 2009}, {\em 79}, 195208.

\bibitem{Analytis2010}
Analytis, James G.; McDonald, Ross D.; Riggs, Scott C; Chu, Jiun-Haw; Boebinger, G.S.; and Fisher, I.R.. Two-dimensional surface state in the quantum limit of a topological insulator. {\em Nat. Phys.} {\bf 2010}, {\em 6}, 960.

\bibitem{Checkelsky2011}
Checkelsky, J.G.; Hor, Y.S.; Cava, R.J.; and Ong, N.P.. Bulk band gap and surface state conduction obsered in voltage-tuned crystals of the topological insulator Bi$_{2}$Se$_{3}$. {\em Phys. Rev. Lett.} {\bf 20011}, {\em 106}, 196801.

\bibitem{Zhong2015}
Zhong, Ruidan; He, Xugang; Schneeloch, J.A.; Zhang, Cheng; Liu, Tiansheng; Pletikosi\'{c}, I.; Yilmaz, T.; Sinkovic, B.; Li, Qiang; Ku, Wei; Valla, T.; Tranquada, J.M.; and Gu, G.D.. Surface-state-dominated transport in crystals of the topological crystalline insulator In-doped \pst. {\em Phys. Rev. B} {\bf 2015}, {\em 91}, 195321.

\bibitem{Springer}
Villars, Pierre. PAULING FILE In \textit{Inorganic Solid Phases}; SpringerMaterials (online database): Springer, Heidelberg, 1998; Volume 41C, pp. 1-8.

\bibitem{Haldo2016}
Haldolaarachchige, Neel; Gibson, Quinn; Xie, Weiwei; Nielsen, Morten Bormann; Kushwaha, Satya; and Cava, R. J.. Anomalous composition dependence of the superconductivity in In-doped SnTe. {\em Phys. Rev. B} {\bf 2016}, {\em 93}, 024520.

\bibitem{Du2015}
Du, Guan; Du, Zengyi; Fang, Delong; Yang, Huan; Zhong, R. D; Schneeloch, J.; Gu, G. D.; and Wen Hai-Hu. Fully gapped superconductivity in In-doped topological crystalline insulator \pstn. {\em Phys. Rev. B} {\bf 2015}, {\em 92}, 020512(R).

\bibitem{Dixon1968}
Dixon, J. R. and Bis, R. F.. Band inversion and the electrical properties of  Pb$_{x}$Sn$_{1-x}$Te. {\em Phys. Rev.} {\bf 1968}, {\em 176}, 942.

\bibitem{Vul1978}
Vul, B. M.; Voronova, I. D.; Kalyuzhnaya, G. A.; Mamedov, T. S.; and Ragimova, T. Sh.. Characteristics of transfer phenomena in Pb$_{0.78}$Sn$_{0.22}$Te with a large indium content. {\em Phys. Rev. Lett.} {\bf 1966}, {\em 26}, 1193.

\bibitem{Kozub2006}
Kozub, V. I.; Parfen'ev, R. V.; Shamshur, D. V.; Shakura, D. V.; Chernyaev, A. V.; and Nemov, S. A.. Superconductor-insulator transition in (Pb$_{1-z}$Sn$_{z}$)$_{0.84}$In$_{0.16}$Te. {\em JETP Lett.} {\bf 1979}, {\em 29}, 18.

\bibitem{Shamshur2008}
Shamshur, D. V.; Nemov, S. A.; Parfen'ev, R. V.; Kononchuk, M. S.; and Nizhankovskii, V. I.. Low-temperature conductivity and the Hall effect in (Pb$_{1-z}$Sn$_{z}$)$_{0.84}$In$_{0.16}$Te semiconducting solid solutions. {\em Phys. Solid State} {\bf 2008}, {\em 50}, 2028.

\bibitem{Bansal2012}
Bansal, Namrata; Kim, Yong Seung; Brahlek, Matthew; Edrey, Eliav; and Oh, Seongshik. Thickness-independent transport channels in topological insulator Bi$_{2}$S3$_{3}$ thin films. {\em Phys. Rev. Lett.} {\bf 2012}, {\em 109}, 116804.

\bibitem{Serbyn2014}
Serbyn Maksym and Fu, Liang. Symmetry breaking and Landau quantization in topological crystalline insulators. {\em Phys. Rev. B} {\bf 2014}, {\em 90}, 035402.

\bibitem{Akiyama2014}
Akiyama, R.; Fujisawa, K.; Sakurai, R.; and Kuroda, S.. Weak antilocalization in (111) thin films of a topological crystalline insulator SnTe. {\em J. Phys.: Conference Series} {\bf 2014}, {\em 568}, 052001.

\bibitem{Kim2011WAL}
Kim, Yong Seung; Brahlek, Matthew; Bansal, Namrata; Edrey, Eliav; Kapilevich, Gary A.; Iida, Keiko; Tanimura, Makoto; Horibe, Yoichi; Cheong, Sang-Wook; and Oh, Seongshik. Thickness-dependent bulk properties and weak antilocalization effect in topological insulator Bi$_{2}$Se$_{3}$. {\em Phys. Rev. B} {\bf 2011}, {\em 84}, 073109.

\bibitem{Hikami1980}
Hikami, S.; Larkin, A. I.; and Nagaoka, Y.. Spin-Orbit Interaction and Magnetoresistance in the Two Dimensional Random System. {\em Prog. Theor. Phys.} {\bf 1980}, {\em 63}, 707.

\bibitem{Assaf2014}
Assaf, B. A.; Katmis, F.; Wei, P.; Satpati, B.; Zhang, Z.; Bennett, S. P.; Harris, V. G.; Moodera, J. S.; and Heiman, D.. Quantum coherent transport in SnTe topological crystalline insulator
thin films. {\em Appl. Phys. Lett.} {\bf 2014}, {\em 105}, 102108.

\bibitem{Kushwaha2016}
Kushwaha, S. K.; Pletikosi\'{c}, I.; Liang, T.; Gyenis, A.; Lapidus, S. H.; Tian, Yao; Zhao, He; Burch, K. S.; Lin, Jingjing; Wang, Wudi; Ji, Huiwen; Fedorov, A. V.; Yazdani, Ali; Ong, N. P.; Valla, T.; and R. J. Cava. Sn-doped Bi$_{1.1}$Sb$_{0.9}$Te$_{2}$S bulk crystal topological insulator with excellent properties. {\em Nat. Commun.} {\bf 2016}, {\em 7}, 11456.

\bibitem{Ravich2002}
Ravich, Yu. I. and Nemov, S. A.. Hopping conduction via strongly localized impurity states of indium in PbTe and its solid solutions. {\em Semiconductors} {\bf 2002}, {\em 36}, 1.

\bibitem{Kaidanov1985}
Ka$\breve{i}$danov, V. I.; and Ravich, Yu. I.. Deep and resonance states in A$^{IV}$B$^{VI}$. {\em Sov. Phys. Usp.} {\bf 1985}, {\em 28}, 31.

\bibitem{Pankratov1987}
Pankratov, O. A.; Pakhomov, S. V.; and Volkov, B. A.. Supersymmetry in heterojuctions: Band-inverting contact on the basis of Pb$_{1-x}$Sn$_{x}$Te and Hg$_{1-x}$Cd$_{x}$Te. {\em Solid State Commun.} {\bf 1987}, {\em 61}, 93-96.

\bibitem{Dimmock1966}
Dimmock, J. O.; Melngailis, I.; and Strauss, A. J.. Band structure and laser action in Pb$_{x}$Sn$_{1-x}$Te. {\em Phys. Rev. Lett.} {\bf 1966}, {\em 26}, 1193.

\bibitem{Gao2008}
Gao, Xing and Daw, Murray S.. Investigation of band inversion in (Pb,Sn)Te alloys using $ab initio$ calculations. {\em Phys. Rev. B} {\bf 2008}, {\em 77}, 033103.

\bibitem{Tanaka2013}
Tanaka, Y.; Sato, T.; Nakayama, K.; Souma, S.; Takahashi, T.; Ren, Z.; Novak, M.; Segawa, K.; and Ando, Y.. Tunability of the $k$-space location of the dirac cones in the topological crystalline insulator Pb$_{1-x}$Sn$_{x}$Te. {\em Phys. Rev. B} {\bf 2013}, {\em 87}, 155105.

\bibitem{Zhang2013_SnTe}
Zhang, Qian; Liao, Bolin; Lan, Yucheng; Lukas, Kevin; Liu, Weishu; Esfarjani, Keivan; Opeil, Cyril; Broido, David; Chen, Gang; and Ren, Zhifeng. High thermoelectric performance by resonant dopant indium in nanostructured SnTe. {\em Proc. Natl. Acad. Sci.} {\bf 2013}, {\em 110}, 13261.

\bibitem{Bushmarina1991}
Bushmarina, G. S.; Drabkin, I. A.; Mashovets, D. V.; Parfeniev, R. V.; Shamshur, D. V.; and Shachov, M. A.. Superconducting properties of the SnTe-PbTe system doped with indium. {\em Physica B: Condens. Matter} {\bf 1991}, {\em 169}, 687.

\bibitem{Heremans2012}
Heremans, Joseph P.; Wiendlocha, Bartlomiej; and Chamoire, M.. Resonant levels in bulk thermoelectric semiconductors. {\em Energy Environ. Sci.} {\bf 2012}, {\em 5}, 5510.

\bibitem{Maeda2017}
Maeda, Satoki; Katsube, Shota; and Zheng, Guo-qing. Quasi-localized impurity state in doped topological crystalline insulator Sn$_{0.9}$In$_{0.1}$Te probed by $^{125}$Te-NMR. {\em J. Phys. Soc. Jpn.} {\bf 2017}, {\em 86}, 024702.

\bibitem{Qi2009}
Qi, Xiao-Liang; Hughes, Taylor L.; Raghu, S.; and Zhang, Shou-Cheng. Time-reversal invariant topological superconductors and superfluids in two and three dimensions. {\em Phys. Rev. Lett.} {\bf 2009}, {\em 102}, 187001.

\bibitem{Qi2010}
Qi, Xiao-Liang; Hughes, Taylor L.; Raghu, S.; and Zhang, Shou-Cheng. Topological invariants for the Fermi surface of a time-reversal-invariant superconductor. {\em Phys. Rev. B} {\bf 2010}, {\em 81}, 134508.

\bibitem{Sasaki2011}
Sasaki, Satoshi; Kriener, M.; Segawa, Kouji; Yada, Keiji; Tanaka, Yukio; Sato, Masatoshi; and Ando, Yoichi. Topological superconductivity in Cu$_{x}$Bi$_{2}$Se$_{3}$. {\em Phys. Rev. Lett.} {\bf 2011}, {\em 107}, 217001.

\bibitem{Kirzhner2012}
Kirzhner, T.; Lahoud, E.; Chaska, K. B.; Salman, Z.; and Kanigel, A.. Point-contact spectroscopy of Cu$_{0.2}$Bi$_{2}$Se$_{3}$ single crystals. {\em Phys. Rev. B} {\bf 2012}, {\em 86}, 064517.

\bibitem{Sasaki2012}
Sasaki, Satoshi; Ren, Zhi; Taskin, A. A.; Segawa, Kouji; Fu, Liang; and Ando, Yoichi. Odd-parity pairing and topological superconductivity in a strongly spin-orbit coupled semiconductor. {\em Phys. Rev. Lett.} {\bf 2012}, {\em 109}, 217004.

\bibitem{Levy2013}
Levy, Niv; Zhang, Tong; Ha, Jeonghoon; Sharifi, Fred; Talin, A. Alec; Kuk, Young; and Stroscio, Joseph A.. Experimental evidence for $s$-wave pairing symmetry in superconducting Cu$_{x}$Bi$_{2}$Se$_{3}$ single crystals using a scanning tunneling microscope. {\em Phys. Rev. Lett.} {\bf 2013}, {\em 110}, 117001.

\bibitem{He2013}
He, L. P.; Zhang, Z.; Pan, J.; Hong, X. C.; Zhou, S. Y.; and Li, S. Y.. Fully superconducting gap in the doped topological crystalline insulator Sn$_{0.6}$In$_{0.4}$Te. {\em Phys. Rev. B} {\bf 2013}, {\em 88}, 014523.

\bibitem{Saghir2014}
Saghir, M.; Barker, J. A. T.; Balakrishnan, G.; Hillier, A. D.; and Lees, M. R.. Superconducting properties of Sn$_{1-x}$In$_{x}$Te ($x$=0.38-0.45) studied using muon-spin spectroscopy. {\em Phys. Rev. B} {\bf 2014}, {\em 90}, 064508.

\bibitem{Matano2016}
Matano, K.; Kriener, M.; Segawa, K.; Ando, Y.; and Zheng, Guo-qing. Spin-rotation symmetry breaking in the superconducting state of Cu$_{x}$Bi$_{2}$Se$_{3}$. {\em Nat. Phys.} {\bf 2016}, {\em 12}, 852.

\bibitem{Arpino2014}
Arpino, K. E.; Wallace, D. C.; Nie, Y. F.; Birol, T.; King, P. D. C.; Chatterjee, S.; Uchida, M.; Koohpayeh, S. M.; Wen, J.-J.; Page, K.; Fennie, C. J.; Shen, K. M.; and McQueen, T. M.. Evidence for topologically protected surface states and a superconducting phase in [Tl$_{4}$](Tl$_{1-x}$Sn$_{x}$)Te$_{3}$. {\em Phys. Rev. Lett.} {\bf 2014}, {\em 112}, 017002.

\bibitem{Arpino2015}
Arpino, K. E.; Wasser, B. D.; and McQueen, T. M.. Superconducting dome and crossover to an insulating state in [Tl$_{4}$]Tl$_{1-x}$Sn$_{x}$Te$_{3}$. {\em APL Mater.} {\bf 2015}, {\em 3}, 041507.

\bibitem{Fu2008}
Fu, Liang and Kane, C. L.. Superconducting proximity effect and Majorana fermions at the surface of a topological insulator. {\em Phys. Rev. Lett.} {\bf 2008}, {\em 100}, 096407.

\bibitem{Wang2012}
Wang, Mei-Xiao; Liu, Canhua; Xu, Jin-Peng; Yang, Fang; Miao, Lin; Yao, Meng-Yu; Gao, C. L.; Shen, Chenyi; Ma, Xucun; Chen, X.; Xu, Zhu-An; Liu, Ying; Zhang, Shou-Cheng; Qian, Dong; Jia, Jin-Feng; and Xue, Qi-Kun. The coexistence of superconductivity and topological order in the Bi$_{2}$Se$_{3}$ thin films. {\em Science} {\bf 2012}, {\em 336}, 52.

\bibitem{Wang2013}
Wang, Eryin; Ding, Hao; Fedorov, Alexei V.; Yao, Wei; Li, Zhi; Lv, Yan-Feng; Zhao, Kun; Zhang, Li-Guo; Xu, Zhijun; Schneeloch, John; Zhong, Ruidan; Ji, Shuai-Hua; Wnag, Lili; He, Ke; Ma, Xucun; Genda, Gu; Yao, Hong; Xue, Qi-Kun; Chen, Xi; and Zhou, Shuyun. Fully gapped topological surface states in Bi$_{2}$Se$_{3}$ films induced by a $d$-wave high-temperature superconductor. {\em Nat. Phys.} {\bf 2013}, {\em 9}, 621.

\bibitem{Xu2014}
Xu, Jin-Peng; Liu, Canhua; Wang, Mei-Xiao; Ge, Jianfeng; Liu, Zhi-Long; Yang, Xiaojun; Chen, Yan; Liu, Ying; Xu, Zhu-An; Gao, Chun-Lei; Qian, Dong; Zhang, Fu-Chun; and Jia, Jin-Feng. Artificial topological superconductor by the proximity effect. {\em Phys. Rev. Lett.} {\bf 2014}, {\em 112}, 217001.

\bibitem{Li2014}
Li, Zheng-zao; Zhang, Fu-Chun; and Wang, Qian-Hua. Majorana modes in a topological insulator/$s$-wave superconductor heterostructure. {\em Sci. Rep.} {\bf 2014}, {\em 4}, 6363.

\bibitem{Xu2015}
Xu, Jin-Peng; Wang, Mei-Xiao; Liu, Zhi Long; Ge, Jian-Feng; Yang, Xiaojun; Liu, Canhua; Xu, Zhu An; Guan, Dandan; Gao, Chun Lei; Qian, Dong; Liu, Ying; Wang, Qiang-Hua; Zhang, Fu-Chun; Xue, Qi-Kun; and Jia, Jin-Feng. Experimental detection of a Majorana mode in the core of a magnetic vortex inside a topological insulator-superconductor Bi$_{2}$Te$_{3}$/NbSe$_{2}$ heterostructure. {\em Phys. Rev. Lett.} {\bf 2015}, {\em 114}, 017001.


\end{thebibliography}
\end{document}